\documentclass[%
 reprint,
 amsmath,amssymb,
 aps,
]{revtex4-2}

\usepackage{graphicx}
\usepackage{dcolumn}
\usepackage{bm}
\usepackage{hyperref}
\usepackage{orcidlink}

\begin{document}


\title{Non-Conservative Mechanics in the Compact-Body Mass-Shell Model and its Waveform Structuring}

\author{Noah M. MacKay~\orcidlink{0000-0001-6625-2321}}
 \email{noah.mackay@uni-potsdam.de}
\affiliation{%
 Institut für Physik und Astronomie, Universität Potsdam\\
 Karl-Liebknecht-Straße 24/25, 14476 Potsdam, Germany
}%

\date{\today}

\begin{abstract}
A novel reinterpretation on the effective one-body (EOB) mapping of compact-body binary (CBB) coalescence was previously considered. Instead of the conventional ``marble in a funnel" description of EOB, CBB coalescence is modeled as the concurrent rotation and contraction of a hollow shell with a constant reduced mass measure. Merger is achieved, therefore, when the shell radius reaches the total mass radius. While past works analyzed the energetics sector of the model and the radiated gravitational waves (GWs), this work considers the geometric waveform modeling aspect. More specifically, we mature the mass-shell waveform analysis initially discussed with toy-model conditions by imposing necessary non-conservative mechanics, leading to the analytical expression of usable mass-shell modeled GW-forms. This is to set the foundation for eventual computational endeavors with the analytical waveforms presented here, enabling a comparison against current and state-of-the-art models.
\end{abstract}

\maketitle


\section{Introduction} \label{intro}

Since their discovery on September 14, 2015, gravitational waves (GWs) have been routinely observed by the LIGO-Virgo-KAGRA (LVK) collaboration \cite{GWOSC, LIGOScientific:2018mvr, LIGOScientific:2021usb, KAGRA:2021vkt, LIGOScientific:2025slb}, now called the International Gravitational Wave Network (IGWN).  The sources of measurable GWs, i.e. the GWs detected across the IGWN, are well understood to be quadrupolar sources, more specifically compact-body binaries (CBBs). Following a linearized gravity regime in the Einstein field equations (EFEs) under a weak-field limit and given a source, one obtains an expression of the traceless-transverse (TT) gauged waveform that is time- and source-dependent:
\begin{equation}\label{gwfunct}
 h^\mathrm{TT}_{ij}=\frac{2}{D}\ddot{Q}_{ij}^\mathrm{TT},
\end{equation}
where $Q_{ij}^\mathrm{TT}$ is the TT-gauged quadrupole that generates the waveforms, and $D$ is the luminosity distance between the observer and the source. Here, we use units where $G=c=1$. It is at this point of waveform modeling where $h^\mathrm{TT}_{ij}$ can hold any form, which is reflected in how one approaches the quadrupole source. E.g.,  if one models the source as a compact binary with two point-mass constituents following simple circular orbits, one obtains via Eq. (\ref{gwfunct})
\begin{equation}\label{gwsourced}
 h^\mathrm{TT}_{ij}=-\frac{4\mu L^2\Omega^2}{D}\exp(i\,2\Omega t)\varepsilon^\mathrm{TT}_{ij},
\end{equation} 
where $\mu=m_1m_2/M$ is the reduced mass of the binary ($M=m_1+m_2$ is the total mass), $L$ is the binary separation, $\Omega$ is the orbital frequency,  $h_+\propto\cos(2\Omega t)$ and $h_\times\propto\sin(2\Omega t)$ via the $3\times3$ TT-gauged matrix $\varepsilon^\mathrm{TT}_{ij}$, and the strain amplitude is distinctly defined in source-dependent values. 

However, realistic CBBs complicate the simple waveforms to incorporate dynamic post-Newtonian (PN) \cite{Blanchet:2013haa} and perturbative post-Minkowskian (PM) \cite{Damour:2016gwp} corrections. These corrections cumulate across early inspiral until reaching the inspiral-merger-ringdown (IMR) process. In these latter phases, the waveform intensifies with a dynamic frequency and amplitude enhancement, until reaching the maximum peak at the coalescence time $t_C$. Past coalescence, ignoring tidal deformations, the waveform dampens exponentially towards a zero flat-line.

One example of an analytical effort to describe CBB dynamics is the effective one-body (EOB) re-mapping, describing the CBB's reduced mass as an inspiraling point particle \cite{Buonanno:1998gg, Buonanno:2005xu} within a deformed background (i.e., adopting a ``marble in a funnel" phenomenology). In this description, the effective reduced-mass particle inspirals from large separations towards the total mass innermost stable circular orbital (ISCO) radius, until it plunges into the total mass horizon with radius $r_S=2M$. The largeness of the symmetric mass ratio $\nu:=\mu/M\in(0,~1/4)$ strongly influences the longevity of nearly-circular orbital paths and how soon they transition into plunge paths \cite{Buonanno:1998gg}. 

The energetics of CBBs in the EOB framework is encoded in the EOB Hamiltonian $H_\mathrm{EOB}$ \cite{Buonanno:1998gg, Damour:2009zoi, Damour:2012mv}, which reads as \cite{Buonanno:1998gg}:
\begin{subequations}
\begin{equation}
H_\mathrm{EOB}=M\sqrt{1+2\nu\left(\frac{H_\mathrm{eff}^\nu}{\mu}-1\right)},
\end{equation}
where
\begin{equation}
H_\mathrm{eff}^\nu=\mu\sqrt{A_\nu(r)\left[1+\frac{\mathbf{p}^2}{\mu^2}+\left(\frac{1}{B_\nu(r)}-1\right)\frac{p_r^2}{\mu^2} \right]}
\end{equation}
with
\begin{equation}
ds^2_\mathrm{eff}=-A_\nu(r)dt^2+B_\nu(r)dr^2+r^2d\Omega^2.
\end{equation}
\end{subequations}
The effective phase-space variables $\mathbf{p}$ and $p_r$ respectively encode tangential and radial motion of the reduced mass, and the metric functions $A_\nu(r)$ and $B_\nu(r)$ incorporate higher-order conservative dynamics as well as Pad\'e resummation effects, in which e.g. $A_\nu(r)$ takes on an analogous PN expansion from the baseline Schwarzschild potential $f(r)=1-2M/r$.

Although the EOB framework can be systematically derived from the EFEs and augmented via PN and PM information, practical applications rely on numerical evaluation and calibration against high-precision numerical-relativity (NR) waveforms. NR-calibrated EOB models include \texttt{SEOBNRv4} \cite{Bohe:2016gbl}, \texttt{v5} \cite{Khalil:2023kep}, and \texttt{v6} \cite{Gamboa:2026jht},  and the tidal deformation extension \texttt{SEOBNRv4T} \cite{Hinderer:2016eia, Steinhoff:2016rfi, Steinhoff:2021dsn} among many more. Nonetheless, the EOB approach depicts the dynamic GW structure with remarkable accuracy, showing the wave evolution of nearly-circular orbital waves in earlier times $t<t_C$ into the chirp rise ($t\sim t_C$) and the ringdown ($t>t_C$) in the IMR process. These regimes are pieced together by quasi-normal modes, e.g. the ringdown piece incorporated the linearized perturbations from stellar collapse and explosions, derived by Gundlach, Price and Pullin \cite{Buonanno:2005xu, Gundlach:1993tp}.

\subsection{The Mass-Shell Model}

Despite its remarkable accuracy and central role in modern GW analysis, evaluating EOB dynamics and associating waveform families remains computationally (and temporary) expensive. This motivates the exploration of complementary, analytically tractable methods to find key measurables -- more specifically the total radiated energy at coalescence -- however at the trade-off of being an approximation. One such alternative is a hollow mass-shell model for CBBs, introduced in Ref. \cite{MacKay:2024qxj} and expanded upon in Ref. \cite{MacKay:2025uyg}.

Rather than explicitly utilizing the ``marble in a funnel" convention of EOB, the binary masses $m_1$ and $m_2$ are distributed as an equatorial ring of reduced-mass measure. Inclination angles $\iota\in[0,~\pi]$ are ``integrated" over by distributing the ring into a hollow shell. Coalescence is instead mapped as the concurrent rotation and contraction of this hollow shell, until the shell reaches its ``innermost" shell radius of $\rho=2M$, consistent with conventional EOB. A visual aid is provided as Figure 1 in Ref. \cite{MacKay:2025uyg}, where the mass-shell representation is compared with an ideal two-body CBB. 

Essentially, this mass-shell representation is applicable for all stages in the inspiral-merger phases. As discussed in Ref. \cite{MacKay:2024qxj} and summarized in Ref. \cite{MacKay:2025uyg}, the effective (and purely phenomenological) model of the CBB as an evolving mass-shell allows one to map the binary's morphology from early inspiral to near-merger. From this morphology, one could extract e.g. low-frequency waveform profiles and the inspiral-phase radiated energy, naming only a couple projects proposed by the analytical initative of the European Einstein Telescope (ET) collaboration \cite{ET:2025xjr}. The ringdown phase is excluded, as it describes the remnant object with final-state parameters (e.g. mass $M_\mathrm{f}$ and spin $\chi_\mathrm{f}$) that is best represented conventionally. However, more informative and relevant stages -- provided the frequency sensitivities of current-generation laser interferometers across the IGWN -- to apply the mass-shell representation rest in the late-inspiral / pre-merger stages. 

The previous works that introduce this mass-shell model \cite{MacKay:2024qxj, MacKay:2025uyg} only used waveform modeling as a means of discussion, with the main focus of these works being on the GW energetics. In summary, the EFEs were approached variationally, in the similar philosophy of the quantum-mechanical variational method. Much like its quantum-mechanical counterpart (an Ansatz wavefunction is inserted in an impossible Hamiltonian to compute its expectation value), an approximated metric Ansatz that effectively describes the source's geometric signature yields effective energetic information via $G_{\mu\nu}=8\pi T_{\mu\nu}$. This is done in an unconventional, inverted way in solving the EFEs: applying the Laplace-Beltrami formulation of the Ricci tensor to a Kerr metric Ansatz, one obtains the corresponding (effective) energy density $T_{00}$ of the CBB mass shell and its surface energy $E=T_{00}V$. This is found to be time-dependent, aligning with well-understood information that time evolution and phase transitions of the CBB drive the chirp rise enhancement of the waveform and with it a peak value in GW energy released. At the time of coalescence $t_C$, the resulting surface energy depends on the reduced mass $\mu$, the symmetric mass ratio $\nu$, and the normalized orbital spin velocity of the CBB, c.f. \cite{MacKay:2025uyg}:
\begin{equation}\label{energy}
E(t_C)\simeq 0.826\nu \mu \left(1-5.276 \beta_C^2 \right).
\end{equation}
The energy given by Eq. (\ref{energy}) becomes zero at the normalized speed ratio $\beta_C\leq0.435$. This is comparable to the rotational velocity at the total mass ISCO radius: $\beta_\mathrm{ISCO}=0.408$, which may serve as an extra constraint in the speed ratio to neutralize the GW energy according to Eq. (\ref{energy}). 

In Eq. (\ref{energy}), if one considers an ideal equal mass binary ($\mu= M/4$ with $\nu\sim1/4$) with a negligible $\beta_C^2\ll1$, then $E(t_C)\approx M/20$, relating the merger-phase emitted energy as roughly five percent of the total CBB mass. In current GW analysis, roughly five percent of the CBB total mass is the anticipated as the peak GW energy released \cite{Dietrich:convo}. For GW150914: the first dectection sourced by binary black holes (BBHs), one finds that the anticipated energy via Eq. (\ref{energy}) is $3.26M_\odot c^2$, which is comparable to the GWTC cataloged value of $3.1^{+0.4}_{-0.4}M_\odot c^2$ \cite{GWOSC, LIGOScientific:2018mvr}. For GW170817, sourced by binary neutron stars (BNSs), one computes $0.140 M_\odot c^2$ for the GW energy via Eq. (\ref{energy}), which is compared to the unconstrained radiated GW energy $E\geq 0.04 M_\odot c^2$ inferred from observation \cite{GWOSC, LIGOScientific:2018mvr}. 

45 select GW events cataloged in GWTC-1 through -4.0 \cite{GWOSC, LIGOScientific:2018mvr, LIGOScientific:2021usb, KAGRA:2021vkt, LIGOScientific:2025slb} were subject to comparison in Ref. \cite{MacKay:2025uyg} between Eq. (\ref{energy}) and observed coalescence energies. These are gauged by the mass-energy difference between the total mass and the remnant mass, yet one may also compare the model prediction with roughly five percent of the total mass-energy. Via statistical goodness-of-fit tests carried out in \cite{MacKay:2025uyg}, the CBB mass-shell model can predict the radiated energy emitted at merger very well, with no net bias.

\subsection{Paper Layout}

Given the initial success of the mass-shell model, it is natural to propose further analysis into the model and attempt to mature it into another CBB mapping to use in future GW analysis. This work addresses the waveform modeling sector, extending it beyond initial toy-model conditions used in \cite{MacKay:2024qxj, MacKay:2025uyg} into realistic settings. More specifically, we approach in Section \ref{noncon}: (i) proper formalisms for a hollow shell moment of inertia, (ii) how rotational kinetic energy balances with a hollow shell's gravitational self-energy -- and how non-conservative mechanisms arise in this approach --, and (iii) how they can be inserted in the wave expressions via Eq. (\ref{gwfunct}). These branching approaches attempt to address e.g. early-inspiral binary separation, near-merger GW frequency, and estimating the timescale to coalescence $t_C$ according to the mass-shell model.

After obtaining the waveforms expected in this mass-shell approach in Section \ref{res}, we yield a symmetric relation with essential phenomenological waveform scalings (i.e., changing mass dependencies from reduced mass to chirp mass via the symmetric ratio $\nu$). In addition, we discuss a heuristic, toy model Ansatz to address time-dependent CBB frequency, and how this approach contributes to waveform profiling and its time-driven evolution. We make concluding statements in Section \ref{concl} and with it what we hope to pursue after this work.

\section{Reintroducing the mass-shell model}\label{model}

For both stable classical binaries and CBBs, the reduced mass $\mu$ simplifies two-body dynamics into a singular effective system. For CBBs specifically, the binary may be assumed to behave as a singular object when viewed from really far away. Rather than taking on the ``marble in a funnel" representation, suppose it takes on the mass-shell representation. While undergoing coalescence, the rate of change in the separation $-\dot{L}$ into a final separation $L(t_C)$ relates to the shell's contracting diameter with the rate of change $-\dot{S}$ and a final measure $S(t_C)$. The overdot resembles a time differential: $^\bullet\equiv d/dt$. When integrated over the timelapse $t'\in[t,\,t_C]$, where $t$ is dynamic and $t_C$ is fixed, we yield:
\begin{equation}\label{seps}
-S(t_C)+S(t)=-L(t_C)+L(t).
\end{equation}
In Eq. (\ref{seps}), the CBB separation and shell diameter at the coalescence time $t_C$ are fixed values. Adopting EOB convention by imposing coalescence ends when the masses make contact: $L(t_C)=r_1+r_2$, under the shell diameter that identifies the total mass horizon diameter: $S(t_C)=4M$, we define:
\begin{equation}\label{rads}
S(t)=L(t)-r_1-r_2+4M.
\end{equation}
Thus, the CBB mass-shell model reinterprets the EOB picture as a shrinking hollow shell with the constant mass measure $\mu$ and contracting radius $\rho(t)=S(t)/2$, until reaching the ``innermost'' shell that is the total mass horizon. 

For a CBB treated as an effective compact object rotating at the dynamic orbital velocity $\Omega=\Omega(t)$,  the quadrupole moment is proportional to the hollow shell moment of inertia $I=2\mu \rho^2/3$. If we assume (nearly-)circular orbits,
\begin{equation}\label{qij}
Q_{ij}=\frac{1}{6}\mu S^2\left(c_ic_j-\frac{1}{3}\delta_{ij}\right),
\end{equation}
with $\vec{c}=(\cos(\Omega t),\,\sin(\Omega t),0)$. One introduces eccentric/elliptical orbits at this stage of the derivation, as one would set e.g. $\vec{c}=2(r_+\cos(\Omega t)+u,\,r_-\sin(\Omega t)+w,0)/S$ to encode the semi-major and semi-minor radii and the center displacement $(u,w)$, or alternatively by allowing a PN-like expansion in $\Omega$ to include an off-circular correction. Nonetheless, via Eq. (\ref{gwfunct}) and using Eq. (\ref{qij}), we yield the (nearly-)circular waveform readily in the TT-gauge, also considering time variation in the CBB separation $S=S(t)$ and angular velocity $\Omega$:
\begin{widetext}
\begin{subequations}
 \begin{equation}\label{hplus}
h_{+}^{\mathrm{TT}}=-\frac{2}{3}\frac{\mu S^2}{D}\Big(\sin(2\Omega t)\left(\frac{2\dot{S}}{S}\left(\Omega +t\dot{\Omega}\right)+\dot{\Omega}+\frac{t}{2}\ddot{\Omega}\right) -\cos(2\Omega t)\left(\frac{\dot{S}^2}{2S^2}+\frac{\ddot{S}}{2S}-\left(\Omega+t\dot{\Omega}\right)^2\right)+\frac{\dot{S}^2}{2S^2}+\frac{\ddot{S}}{2S}\Big),
\end{equation}
\begin{equation} \label{hcross}
h_{\times}^\mathrm{TT}=-\frac{2}{3}\frac{\mu S^2}{D}\Big(\sin(2\Omega t)\left(\frac{\dot{S}^2}{2S^2}+\frac{\ddot{S}}{2S}-\left(\Omega+t\dot{\Omega}\right)^2\right)+\cos(2\Omega t)\left(\frac{2\dot{S}}{S}\left(\Omega +t\dot{\Omega}\right)+\dot{\Omega} +\frac{t}{2}\ddot{\Omega}\right)\Big).
\end{equation}
\end{subequations}
\end{widetext}

If we assume, for early inspiral stages, large separations such that $S\approx L$ and perfectly circular orbits, where all rates of change are zero, we recover the wave profiles nearly identical to Eq. (\ref{gwsourced}), however scaled by $1/6$ due to our choice of quadrupole geometry. To restore the factor of $4$, one can instead adopt the standard two-point-mass quadrupole moment, which bypasses the moment of inertia in favor of $m_1r_1^2+m_2r_2^2=\mu L^2$. In this case, if one chooses not to approximate $S\approx L$ and keep $S$ in the waveforms, one would then rewrite $L$ by rearranging Eq. (\ref{rads}). This ultimately becomes a discrepancy in the scaling factor within the order of unity; one may argue that the scaling factor within the order of unity is irrelevant, as long as the structuring of essential parameters that compute the strain amplitude and its essential order of magnitude of $h\in10^{-22}\sim10^{-21}$ remain unchanged.

However, unlike the simple circular wave model in which both separation and angular frequency are fixed, these quantities are dynamic across coalescence. With orbital mechanics of the CBB now described as effective axial rotations, the mass-shell model has an angular momentum $J=I\Omega$. Simplistic toy model discussions in \cite{MacKay:2024qxj, MacKay:2025uyg} considered a conservation in $J$, i.e. $\dot{J}=0$, which enables a geometric relation whereby $\dot{\Omega}\propto-\dot{S}$, showing that the decrease in CBB separation $-\dot{L}=-\dot{S}$ leads to an increase in orbital (mass-shell axial) rotation. While this might be intuitive to describe the nature of coalescence, imposing the conservation in angular momentum suggests that no GW information had been radiated out. It is after all the loss in binding energy of the CBB that contributes to the radiation of GWs; the internal mechanisms are inherently non-conservative. Therefore, we must impose in this study $\dot{J}\neq0$, and how this impacts the mass-shell waveform expressions containing $S,~\dot{S},~\ddot{S}$ and $\Omega,~\dot{\Omega},~\ddot{\Omega}$. 

\subsection{Inserting Non-Conservative Mechanisms}\label{noncon}

Even in a pedagogical Newtonian approach, a gravitationally bound orbital system is expected to emit energy. Briefly considering the ``marble in a funnel" phenomenology of EOB in the Newtonian context, the inspiraling particle of reduced-mass measure $\mu$ has a centripetal acceleration that is equal to its gravitational attraction to an effective total mass central object. With $v_\perp^2=M/r$ as a result of this, where $r$ is the orbital separation, then the total (the kinetic-plus-potential) energy is non-trivial:
\begin{equation}
E=-\frac{1}{2}\frac{\mu M}{r}.
\end{equation}
If we impose inspiral in this scenario, one finds that $\dot{E}/E= -\dot{r}/r$. In other words, the growing release of binding energy is dependent on the shrinkage in the seperation. In the EOB context, once $r=2M$ to reach merger, the radiated energy obtained from our above expression is $E_\text{rad}=-\mu/4=-\nu M/4$. Of course, the radiated energy propagated as GWs is $\sim M/20$, and this simplistic Newtonian model slightly overestimates this energy with $\sim M/16$ for $\nu\sim1/4$. One must, therefore, consider the essential post-Newtonian corrections and a proper general-relativistic approach. We reassure that this back-of-the-envelope calculation is only a proof of concept, demonstrating that the mechanics of any inspiraling system is inherently non-conservative.

In the mass-shell representation, the approach is different. Here, the CBB behaves both like a rotating hollow shell and a gravitating body. In other words, the hollow shell  has both a rotational kinetic energy $T=I\Omega^2/2$ (which one can rewrite as $T=J\Omega/2$) and a gravitational self energy $U=\mu^2/(2\rho)$, where $2\rho=S$. Keeping $U$ as implicit and defining the equivalence $T=U$, we motivate time-dependent changes in both forms of energy by imposing time derivatives on $J\Omega/2$ and $U\propto S^{-1}$:
\begin{equation}\label{selfs}
 \dot{\left(J\Omega\right)} = 2\dot{U}\implies \dot{J}=2\frac{\dot{U}}{\Omega}-J\frac{\dot{\Omega}}{\Omega}.
\end{equation}
Defining the following from the hollow shell self energy:
\begin{equation}\label{us}
\dot{U}=-\frac{\mu^2}{S^2}\dot{S}=-U\frac{\dot{S}}{S},
\end{equation}
$\dot{J}$ in Eq. (\ref{selfs}) is revised as follows, after factoring out $J$ on the right-hand side:
\begin{equation}
\dot{J}=-J\left(\frac{2U}{J\Omega}\frac{\dot{S}}{S}+\frac{\dot{\Omega}}{\Omega}\right).
\end{equation}
This shows, here, that $\dot{J}\propto-J$, and given the initial equivalence $T=U$ with $T=J\Omega/2$, $2U/(J\Omega)=1$.

 Now considering the geometric sector of $\dot{J}$, where $J=2\mu\rho^2\Omega/3$ for a hollow shell with $\rho=S/2$, we impose explicit time dependence on $\rho\propto S$ and $\Omega$ while keeping $\dot{J}\neq0$ implicit:
\begin{equation}
\dot{J}=\frac{1}{3}\mu S\left(\dot{S}\Omega+\frac{1}{2}S\dot{\Omega} \right).
\end{equation}
After one factors out $J=\mu S^2\Omega/6$ to define $\dot{J}/J$, we form an equivalence between the geometric and energetic sectors:
\begin{equation}\label{equiv}
\frac{\dot{J}}{J}=\left(2\frac{\dot{S}}{S}+\frac{\dot{\Omega}}{\Omega} \right)\equiv -\left(\frac{\dot{S}}{S}+\frac{\dot{\Omega}}{\Omega}\right),
\end{equation}
and by combining like terms we come to the final equivalence linking the change in shell diameter with its axial frequency:
\begin{equation}\label{mash}
\implies -\frac{\dot{S}}{S} = \frac{2}{3}\frac{\dot{\Omega}}{\Omega}.
\end{equation}
From this, we also define $\ddot{S}/S$ that appears in both Eqs (\ref{hplus}) and (\ref{hcross}):
\begin{equation}\label{mash2}
 \frac{\ddot{S}}{S} = -\frac{2}{3}\left(\frac{\ddot{\Omega}}{\Omega}-\frac{\dot{\Omega}^2}{\Omega^2} \right)+\frac{\dot{S}^2}{S^2},
\end{equation}
where the mass-shell waveforms can be written fully in terms of the rates of change in e.g. $\Omega$.

\subsubsection{Early-Inspiral Binary Separation}

Eq. (\ref{mash}) is different from the toy-model condition $-2\dot{S}/S=\dot{\Omega}/{\Omega}$ from  \cite{MacKay:2024qxj}, which comes from the conserved mechanics $\dot{J}=0$ in the geometric sector. As a result of imposing non-conservative mechanics and allowing dissipation in the form of $\dot{J}\propto-J$ in the energetics sector, this changes the scaling in the relation. And as it is a logarithmic-ratio ODE, this eventually influences the power relation between the two quantities. For Eq. (\ref{mash}), we can solve this ODE via integration from the initial and final states in both $S$ and $\Omega$ to form the expression
\begin{equation}
\log\left(\frac{S_0}{S_f} \right)=\frac{2}{3}\log\left(\frac{\Omega_f}{\Omega_0}\right).
\end{equation} 
Lifting the logarithm via the exponential, we yield $S\propto\Omega^{-2/3}$. This recovers the essential Keplerian relation for orbital mechanics $\Omega^2\propto S^{-3}$.

From this, we insert that $S_0\simeq L_0$ for an initial shell diameter to be as far as an initial separation, and $S_f=4M$ for the total mass horizon diameter. In addition we recall that the mass-shell's axial frequency is the CBB orbital frequency, which in turn scales with the GW frequency $f_\mathrm{GW}$. Thus, the final orbital frequency relates to the maximum GW frequency $f_\mathrm{GW,max}$ and the initial orbital frequency to the minimum GW frequency $f_\mathrm{GW,min}$. We obtain an expression of the length scale of $L_0/M$, and with it an expression for the initial separation:
\begin{equation}\label{l0}
\implies {L_0} \simeq 4M\left(\frac{f_\mathrm{GW,max}}{f_\mathrm{GW,min}}\right)^{2/3}.
\end{equation} 
Defining $L_0$ is inherently an order-of-magnitude discussion. Given the above expression, defining $L_0$ strongly depends on how one sets the minimum GW frequency (from which frequency we track the progress of inspiral) and the maximum GW frequency (just before merger). 

Here, we initially set $f_\mathrm{GW,min}=20$ Hz, stating that above this frequency we keep track of CBB inspiral up to merger. The maximum GW frequency is discernible from the \textit{GWOSC} open-access catalog \cite{GWOSC}, and it is unique for each GW event. For the case of BBHs, coalescence frequencies range from $\sim100$ Hz (GW190521) to $\sim250$ Hz (GW150914); this produces a range in $L_0$ to be $12M\sim22M$. For the case of the BNS event GW170817 with coalescence frequency $\sim300$ Hz, we yield $L_0\simeq24M$. 

As it is well understood that smaller-mass CBBs have a longer inspiral, covering more orbits than their larger mass counterparts before reaching merger, we form an intuition that accessibility to higher coalescence frequencies is possible if the scale $L_0/M$ is large. We remind again at this is as long as we manually set $f_\mathrm{GW,min}$, chosen here to be 20 Hz. On the other hand, choosing an even smaller $f_\mathrm{GW,min}$, one would find, further extends $L_0/M$ to higher values. Otherwise, choosing a larger $f_\mathrm{GW,min}$ contracts $L_0/M$ to smaller values.

Our point of reference is the ratio $L/M$ solved in the worldline quantum field theory (WQFT) framework: an effective field theory applicible to early-inspiral gravitational scattering \cite{Mogull:2020sak}. WQFT solves for $L/M$ as an effective impact parameter from elastic scattering amplitudes, and one finds that agreement between WQFT and NR is mutual for binary separations $L>14M$ \cite{Driesse:2024feo}. In this work, we offer separation scales where the ratio $L/M$ is explicitly greater than 14, except for $L_0\simeq12M$ implied by GW190521, and we furthermore offer a maximum constraint to the length scale dependent on CBB type. These are meant to provide supplementary computations to complement the WQFT calculations.

\subsubsection{Near-Merger GW Frequency}

For events in which $f_\mathrm{GW,max}$ is neither reported in the detection paper nor cleanly discernible from the chirp rise in the strain plots within \textit{GWOSC} \cite{GWOSC} (i.e., LVT151012 from GWTC-1 \cite{LIGOScientific:2018mvr} and both GW190403\_051519 and GW190917\_114630 from GWTC-2.1 \cite{LIGOScientific:2021usb}) -- while having mass measurements nonetheless --, the initial analysis of these events in Ref. \cite{MacKay:2025uyg} only considered the leading Newtonian contribution in Eq. (\ref{energy}). As it is known, this is an incomplete calculation, but one can prove that it suffices for least-massive systems, c.f. \cite{MacKay:2025uyg}. 

However, it is ill-advised to assume, effectively, $\beta_C^2\ll1$ even for events that have substantial mass measurements yet have illegible strain plots. This is because $\beta_C= Mf_\mathrm{GW,max}$, and larger mass systems would have a notable spin contribution in the coalescence energy. This would lead to incomplete calculations where the spin information is not accounted for, and thus not reflected in the approximated coalescence energy if naively omitted. Therefore, it is rather useful to derive a ``rule-of-thumb" peak frequency expression from first principles, to use in such cases. 

We once more utilize the detail in the CBB mass-shell model that the orbital frequency of the CBB (i.e., the frequency of radiated GWs) is the axial rotation of the hollow shell acting as a gravitating body. Equating once more the hollow shell rotational kinetic energy to its gravitational self-energy, we yield
\begin{equation}\label{kep}
    \frac{1}{3}\mu\rho^2\Omega^2 = \frac{1}{2}\frac{\mu^2}{\rho}\implies \Omega^2\rho^3=\frac{3}{2}\mu.
\end{equation}
We recover a modified Keplerian limit to the CBB mass-shell's axial rotation, where conventionally for a gravitational mass ${M}_g$ with surface radius $R$: $\Omega^2_K={M}_g/R^3$. Essentially for binaries, $M_g\rightarrow M$ to account for the total mass. 

From Eq. (\ref{kep}), one can define the effective axial rotation for all phases across coalescence given the morphology in $\rho$. For the merger phase at time $t=t_C$, where $\rho=2M$,
\begin{equation}
    \Omega_C=\frac{1}{2M}\sqrt{\frac{3}{4}\frac{\mu}{M}}=\frac{\sqrt{3\nu}}{4M}.
\end{equation}
We recall that the axial rotation of the CBB mass-shell model is the orbital frequency of the CBB, and that the peak GW frequency is one-half the peak orbital frequency. Thus, defining $\Omega_C=\pi f_\text{GW,peak}$, we furthermore define
\begin{equation} \label{freq}
    f_\text{GW,max}=\frac{\sqrt{3\nu}}{4\pi M}.
\end{equation}
Using GW150914 as a representative example, the cataloged peak GW frequency was $\sim250$ Hz. Using the central binary mass values to compute $f_\text{GW,max}$ via Eq. (\ref{freq}), we obtain $216$ Hz from our derived equation provided $M\simeq 64.5M_\odot$ \cite{GWOSC} and $\nu\simeq0.249$. With a 1:1 ratio being $0.844$, this demonstrates a modestly good approximation to what's been inferred from observation. 

However, we must acknowledge that this is more useful for CBBs with component masses of order $10\sim10^2M_\odot$. If we consider GW170817 as a second example, which has the cataloged frequency of $\sim300$ Hz, we obtain a severely overestimated $5.12$ kHz from our expression provided $M\simeq2.73M_\odot$ and $\nu\simeq0.249$. We remind that the BNS source has component masses each of order $\sim1M_\odot$, which allows the caveat that only the Newtonian part of the energy be calculated, being a low-mass system. As an added analytical ensurance, one may introduce the Heaviside function $\Theta(M-10M_\odot)$ to Eq. (\ref{freq}), such that total masses of $M>10M_\odot$ are only considered.

\subsection{Estimating $t_C$ for a Mass-Shell}

From the balance equation between the GW luminosity $\mathcal{L}_\mathrm{GW}$ and the rate of loss of virialized binding energy (in terms of the PN variable $x$) $\mu\dot{x}/2$, one finds an expression for the time to merger $\Delta t_C:=t_C-t$, at leading Newtonian order, in terms of the dynamic binary separation $L(t)$ and the source masses: 
\begin{equation}\label{tc}
\Delta t_C=\frac{5}{256}\frac{L(t)^4}{m_1m_2M}. 
\end{equation}
One can use this expression at the start time $t=0$, where $\Delta t_C=t_C$ and $L(0)=L_0$ (which may be expressed as Eq. (\ref{l0}), to quantify $t_C$ and compare the result with known lapse times up till merger from $f_\mathrm{GW,min}=20$ Hz). 

On the other hand, if one were to express the timelapse purely in terms of the geometry of the contracting mass-shell, we may draw an analogy with cosmology and the Hubble parameter. Rather than describing expansion on cosmological scales, we consider the local contraction of the mass shell. When this contraction admits a self-similar description, the binary separation that defines the mass-shell radius may play a role analogous to a cosmological scale factor. This enables the introduction of a local Hubble-like parameter for the CBB mass-shell, and with it its inverse magnitude as a characteristic timescale. 

Using the Hubble relation $\vec{u}=H\vec{r}$ \cite{Hubble:1929ig}, our current interpretation of $H$ will parameterize the evolution of the contracting hollow shell geometry. And so, the radius vector $\vec{r}$, representing the mass-shell's scale, is modeled after $\rho(t)=S(t)/2$ via Eq. (\ref{rads}) and thus defined as $\vec{r}=\rho(t)\hat{r}$. The radial velocity $\vec{u}$ of the contracting mass-shell is modeled after the inward-pointing radial velocity of the CBB itself. Throughout coalescence, this velocity may be expressed as $\vec{u}=-2\beta^5(M/P)^{1/2}\hat{r}$ (c.f. Ref. \cite{Loutrel:2018ssg}), where $\beta\lesssim0.4$ via Eq. (\ref{energy}) is the normalized rotational speed ratio. Here, $\beta^5$ serves as an effective parameterization of the osculating eccentricity, reflecting the tendency for eccentricity to increase dynamically as coalescence proceeds. Correspondingly, the semi-latus rectum $P$ is roughly $6M$ for nearly circular orbits and $(10\sim15)M$ for orbits with higher eccentricity \cite{Loutrel:2018ssg}.

As discussed in the start of Section \ref{model}, the concurrent rotation and contraction of the CBB mass-shell evolve with time $t$ up to merger. Consequently, the radial velocity $\vec{u}\propto\beta^5 P^{-1/2}$ explicitly becomes time-dependent through the evolution of $\beta(t)$ and the dimensionless semi-latus rectum $\widetilde{P}(t)=P(t)/M$. We therefore define a time-dependent Hubble-like parameter for the CBB mass-shell surface:
\begin{equation} \label{hubble}
H(t)=-\frac{2\beta(t)^5}{\rho(t)\sqrt{\widetilde{P}(t)}}.
\end{equation}
This quantity is negative (due to the negativity of $\vec{u}$), reflecting the contraction of the shell during coalescence. Therefore, its inverse magnitude $\tau=1/|H(t)|$ provides an instantaneous characteristic timescale of mass-shell contraction. And so, for a coalescing CBB, $\tau$ may provide an order-of-magnitude estimate for the associated coalescence lapsetime from a given observational start time $t=t_0$, which may be inferred from initial conditions of the observable parameters. 

Suppose observation begins at the start of inspiral (at $t_0=0$) where $\beta\sim0.1$ and $\widetilde{P}\approx6$. The time to merger at the start of inspiral, at leading Newtonian order, is also given as Eq. (\ref{tc}). Using Eq. (\ref{tc}) to solve for $L(0)\approx 2\rho(0)$ in terms of the masses and $\Delta t_C=t_C$, we define the mass-shell radius to substitute in place of $\rho(0)$ in Eq. (\ref{hubble}) while imposing the timescale condition $\tau=t_C$. From this, we yield an inspiral timelapse expression:
\begin{equation}\label{ins}
t_C^{3/4}=\left(\frac{16}{5} \right)^{1/4}\frac{\sqrt{\widetilde{P}}}{2\beta^5}\left(m_1m_2M \right)^{1/4},
\end{equation}
where $t_C$ via Eq. (\ref{ins}) is isolated by imposing the $4/3$ power on both sides of the equation. 

As an illustrative application, let us consider the BNS source for GW170817 \cite{LIGOScientific:2017vwq}. Above 20 Hz, the GW signal contained a 100-second long inspiral before the BNS coalesced. The properties of the BNS source include, but are not limited to, the binary masses  $m_1=1.46^{+0.12}_{-0.10}~M_\odot$ and $m_2=1.27^{+0.09}_{-0.09}~M_\odot$ \cite{GWOS:170817}, whose central (best fit) values compute the total mass to be $M=2.73~M_\odot$. For our values for $\beta\sim0.1$ and $\widetilde{P}\approx6$ and our BNS information, we yield a model-dependent inspiral timelapse of approximately 383.267, equivalently 6 minutes and 23 seconds. 

This timescale exceeds the recorded 100 seconds above 20 Hz, which we remind is given the current sensitivities of operating IGWN detectors. This suggests that the remaining time of $\sim283$ seconds corresponds to the inspiral lapse time \textit{below} 20 Hz. And using the equations derived in this work so far, we can compute a theoretical expectation of the minimum GW frequency (when inspiral begins) specifically for GW170817, and in principle for other GW transient events.

With this calculation of $t_C$, we can use it in its place in Eq. (\ref{tc}) to find $L(0)\simeq L_0$, and in turn use this to compute $f_\mathrm{GW,min}$ in Eq. (\ref{l0}) provided $M=2.73~M_\odot$ and $f_\mathrm{GW,max}\sim300$ Hz that is particular to GW170817. Following this strategy, we yield $L_0\simeq377.203M_\odot$ (equivalently 556.375 km), and therewith $f_\mathrm{GW,min}=1.478$ Hz. Thus, within the assumptions of the model, the inspiral is predicted to extend to frequencies of order $\sim1$ Hz. Inspiral frequencies in this proximity is below the sensitivity of current-generation GW detectors, and proposed future-generation detectors such as ET aim to capture these low inspiral-phase GW frequencies in the waveform. 

\section{Results} \label{res}

Using Eqs. (\ref{mash}), (\ref{mash2}) and (\ref{kep}), the mass-shell waveforms can be expressed fully in terms of e.g. rates of change in $\Omega$:

\begin{widetext}
\begin{subequations}
 \begin{equation}\nonumber
h_{+}^{\mathrm{TT}}=-\frac{4}{D}\sqrt[3]{\frac{2}{3}}{\mu^{5/3}\Omega^{-4/3}}\left(\sin(2\Omega t)\left(-\frac{4}{3}\dot{\Omega}\left(1 +t\frac{\dot{\Omega}}{\Omega}\right)+\dot{\Omega}+\frac{t}{2}\ddot{\Omega}\right) -\cos(2\Omega t)\Bigg(\frac{7}{9}\frac{\dot{\Omega}^2}{\Omega^2}-\frac{1}{3}\frac{\ddot{\Omega}}{\Omega}-\left(\Omega+t\dot{\Omega}\right)^2\right)
\end{equation}
 \begin{equation}\label{hp}
\quad\quad\quad\quad\quad\quad\quad\quad\quad\quad\quad\quad\quad\quad\quad\quad\quad\quad\quad\quad\quad
\quad\quad\quad\quad\quad\quad\quad\quad\quad\quad+\frac{7}{9}\frac{\dot{\Omega}^2}{\Omega^2}-\frac{1}{3}\frac{\ddot{\Omega}}{\Omega}\Bigg),
\end{equation}
\begin{equation} \label{hc}
h_{\times}^\mathrm{TT}=-\frac{4}{D}\sqrt[3]{\frac{2}{3}}\mu^{5/3}\Omega^{-4/3} \Bigg(\sin(2\Omega t)\left(\frac{7}{9}\frac{\dot{\Omega}^2}{\Omega^2}-\frac{1}{3}\frac{\ddot{\Omega}}{\Omega}-\left(\Omega+t\dot{\Omega}\right)^2\right)+\cos(2\Omega t)\left(-\frac{4}{3}\dot{\Omega}\left(1 +t\frac{\dot{\Omega}}{\Omega}\right)+\dot{\Omega} +\frac{t}{2}\ddot{\Omega}\right)\Bigg).
\end{equation}
\end{subequations}
\end{widetext}

Considering once more the case of perfectly circular orbits, where all rates of change in the orbital frequency are zero, we yield the propotionality of the wave amplitude to be $h^\mathrm{TT}\propto \mu^{5/3}\Omega^{2/3}\sim\mu^{5/3}f^{2/3}$. In previous phenomenological waveform modeling where one instead considers two point-like compact masses, the mass scaling in the frequency-dependent amplitude is set by the chirp mass: $h^\mathrm{TT}\propto\mathcal{M}^{5/3}f^{2/3}$. This is sourced by the Keplerian law that $L^2\propto (M\Omega^{-2})^{2/3}$ for binaries, and $\mu M^{2/3}=\mathcal{M}^{5/3}$. Recognizing that $\mathcal{M}=\nu^{8/5}\mu$, one can correlate the previous (circular-orbit) phenomenological waves with the current derived mass-shell waveforms. 

From this, a symmetric relation arises between the two models, where the normalization factor depends on the symmetric mass ratio:
\begin{equation}\label{symms}
h^\mathrm{TT}_{+/\times,\,\mathrm{shell}}=\frac{1}{\nu^{8/3}}\sqrt[3]{\frac{2}{3}}h^\mathrm{TT}_{+/\times,\,\mathrm{phenom}}.
\end{equation}
As $\sqrt[3]{2/3}\simeq0.873$, one can crudely approximate this as unity, thus presenting $\nu^{-8/3}$ as the essential symmetric normalization between the present mass-shell and previous phenomenological models. Provided $\nu\sim1/4$ in conceivable cases of GW detection, $\nu^{-8/3}\sim4^{8/3}\simeq40.32$. Therefore, given our numerical scalings, one can anticipate that $h^\mathrm{TT}_{\mathrm{shell}}\approx (35\sim40) h^\mathrm{TT}_{\mathrm{phenom}}$. 

This furthermore suggests that the mass-shell waveforms would overestimate the phenomenological strain amplitude at least by one order of magnitude, which would project peak strain values to be of order $10^{-21}\sim10^{-20}$ rather than $10^{-22}\sim10^{-21}$ (the minimum value relates to BNSs, and the maximum value to BBHs). In this regard the phenomenological waveform should be used. Nonetheless, considering that the mass-shell waveforms readily contain the information on the rates of change in $\Omega\sim f$, one may claim that the presented waveform profiles also define the phenomological waveforms in proxy via Eq. (\ref{symms}). 

If one accepts this claim, this work presents additional contributing factors of $\dot{\Omega}$ and $\ddot{\Omega}$ in the time-domain, complementing the well-known information that the frequency changes in a non-linear fashion across inspiral up to merger, which in turn alters the shape of the GWs. 

\subsection{Heuristic Time-Dependent $\Omega,~\dot{\Omega},~\ddot{\Omega}$}

As a heuristic example, we choose the Ansatz $\ddot{\Omega}=\epsilon$, which intends to define a constant ``acceleration" in the orbital frequency that drives its time-dependent, non-linear evolution. In this case, defining $\dot{\Omega}$ and $\Omega$ is done via integration from the initial ``acceleration" gauge:
\begin{subequations}
\begin{equation}\label{omd}
\dot{\Omega}=\epsilon t+\omega,
\end{equation}
\begin{equation}\label{om}
\Omega=\frac{1}{2}\epsilon t^2+\omega t+\Omega_0.
\end{equation}
\end{subequations}
One can see in Eq. (\ref{om}) that the constant term $\Omega_0$ serves as an initial, circular-orbital frequency at $t=0$, and the quadratic term $\epsilon t^2/2$ serves as the main contributor to the quadratic-in-behavior chirp rise in the frequency as $t\rightarrow t_C$. 

With three tunable gauge parameters $\epsilon$, $\omega$ and $\Omega_0$, one would calibrate these values to match a specific event's frequency-time plot inferred in the open-access \textit{GWOSC} catalog \cite{GWOSC}. On the other hand, one should keep in mind that the orbital phase $\Psi=2\Omega t$ of the binary orbit and the GW are conventionally PN-expanded, following the leading orders in $v^2\sim (\Omega M)^{2/3}:=x$ in the frequency domain $\sim f^{2/3}$ \cite{Blanchet:2013haa}. Here, in Eq. (\ref{om}), an expansion arises in leading orders of the dynamic time $t$, i.e. $\Psi= a_3t^3+a_2t^2+a_1t$ structurally. And so, one may expect \textit{a priori} that $\epsilon$, $\omega$, and $\Omega_0$ -- in this toy-model example -- could depend on essential source binary parameters (e.g. eccentricity, tidal deformability, and spin-orbit/-spin coupling). In this case, instead of being coupled to terms of order $x^n$ and being classified as $n$PN structures in the frequency domain, dynamic contributions would be coupled to terms of order $t^\alpha$ and be classified heuristically as ``$\alpha$"PN structures in the time domain. This is beyond the scope of this work, but this would be in adjunct to the PN discussion within the mass-shell model energetics provided in Ref. \cite{MacKay:2025uyg}.

As a leading example we explore the asymptotic behavior of $\omega\rightarrow0$ for a negligible linear contribution, to see how the waveforms change from initial circular waves to final-state waves with a non-linear frequency enhancement. Given these expressions for the rates of change in $\Omega$, and for $\Omega$ itself, the polarization envelope functions $\mathfrak{E}_{+/\times}$ that are specific to Eqs. (\ref{hp}) and (\ref{hc}) are asymptotically defined as follows:
\begin{widetext}
\begin{subequations}
 \begin{equation}
\mathfrak{E}^\mathrm{sin}_+=\mathfrak{E}^\text{cos}_\times\simeq\left(-\frac{4}{3}\epsilon t\left(1 +\frac{2\epsilon t^2}{\epsilon t^2+2\Omega_0}\right)+\frac{3}{2}\epsilon t\right),
\end{equation}
\begin{equation}
\mathfrak{E}^\mathrm{sin}_\times=-\mathfrak{E}^\mathrm{cos}_+\simeq\left(\frac{28}{9}\frac{\epsilon^2 t^2}{(\epsilon t^2+2\Omega_0)^2}-\frac{2}{3}\frac{\epsilon }{\epsilon t^2+2\Omega_0}-\left(\frac{3}{2}\epsilon t^2+\Omega_0\right)^2\right).
\end{equation}
\end{subequations}
\end{widetext}
In addition, the sinusoidal profiles $h_{+/\times}\propto\exp(i\,2\Omega t)$ are revised such that the argument is the polynomial in $t$: $2\Omega t = \epsilon t^3+2\Omega_0 t$. The mass-shell wave amplitude $h\sim\mu^{5/3}\Omega^{-4/3}$ is revised as
\begin{equation}
h\simeq-\frac{4}{D}\sqrt[3]{\frac{2}{3}}\mu^{5/3}\left(\frac{1}{2}\epsilon t^2+\Omega_0\right)^{-4/3}.
\end{equation}
Under this ``very small $\omega$" limit, we consider asymptotic expressions for an infinite time range $t\in[0,~\infty)$, located at the lower and upper time bounds. At $t=0$ (i.e., at the start of measurement), both waveform polarizations have existing cosine and time-independent contributions:
\begin{subequations}
 \begin{equation}
h_{+}^{\mathrm{TT}}(0)=-\frac{4}{D}\sqrt[3]{\frac{2}{3}}\mu^{5/3}\Omega_0^{2/3},
\end{equation}
\begin{equation} 
h_{\times}^\mathrm{TT}(0)=0;
\end{equation}
\end{subequations}
as time progresses towards $t\rightarrow\infty$, the quadratic contribution in $\Omega$ take precidence in both wave envelope and internal sinusoidal profiles:
\begin{subequations}
 \begin{equation}
h_{+}^{\mathrm{TT}}(t\rightarrow\infty)\simeq \frac{18}{D}\left(\frac{\mu^5\epsilon^2t^4}{3}\right)^{1/3}\cos(\epsilon t^3) ,
\end{equation}
\begin{equation}
h_{\times}^\mathrm{TT}(t\rightarrow\infty)\simeq \frac{18}{D}\left(\frac{\mu^5\epsilon^2t^4}{3}\right)^{1/3}\sin(\epsilon t^3).
\end{equation}
\end{subequations}
Even under these upper and lower bounds in time progression, $h_+\propto\cos(2\Omega t)$ and $h_\times\propto\sin(2\Omega t)$ consistently. Between these waveform expressions, one can illustrate a characteristic ``Gabriel's Horn" structure from the initial circular wave profile to the final-state non-linear profiles, while over time the wavelengths tighten. If these cases are mutually interplayed within a more defined time-domain of $t\in[0,~t_C]$, then these mechanisms could draw the quintessential inspiral profile up to merger. 

\subsubsection{Alternative definition of $t_C$}

Briefly recalling the logarithmic ODE in the angular momentum via Eq. (\ref{equiv}), Eq. (\ref{mash}) and the toy-model Ansatz expressions for $\Omega$ and $\dot{\Omega}$ yield
\begin{equation}\label{jt}
\frac{\dot{J}}{J}=-\frac{1}{3}\frac{\dot{\Omega}}{\Omega}=-\frac{2}{3}\frac{\epsilon t}{\epsilon t^2+2\Omega_0}.
\end{equation}
The time-dependent expression for $\dot{\Omega}/\Omega$ reveals insightful information. At the infinite-range time bounds of $t=0$ and $t\rightarrow\infty$, both cases collapse the expression to zero, insinuating \textit{conservative} mechanics (i.e., circular orbits) at the start of inspiral and at infinite time. The expression for $\dot{\Omega}/\Omega$ via Eq. (\ref{jt}) has a quantifiable peak within $t\in[0,~\infty)$, and for a coalescence event that occurs within $t\in[0,~t_C]$ the peak release of GWs occurs at $t\sim t_C$.  Suggesting that this peak value of $\dot{\Omega}/\Omega$ takes place at $t=t_C$, we minimize $\dot{\Omega}/\Omega$ with respect to $t$ at the merger time to define $t_C$:
\begin{equation}
\partial_t\left[\frac{\dot{\Omega}}{\Omega}\right]_{t=t_C}=0\implies t_C=\sqrt{\frac{2\Omega_0}{\epsilon}}.
\end{equation} 
As a proof of concept, if $\epsilon=0$ to insinuate that the CBB is strictly in circular orbits, $t_C\rightarrow\infty$ and coalescence of perfectly circular binaries will never be achieved. Therefore, $\epsilon$ has to be non-zero for coalescence to occur, and how $t_C$ is presently expressed depends on the strength in the toy-model parameters $\epsilon$ and $\Omega_0$. 

And so, if $t=0$ shows mutually $\dot{J}/J=0$ and $\dot{\Omega}/\Omega=0$ for an initially circular CBB, the non-conservative relation for the angular momentum at $t=t_C$ reads as
\begin{equation}
\left.\frac{\dot{J}}{J}\right|_{t=t_C}=-\frac{1}{6}\sqrt{\frac{2\epsilon}{\Omega_0}}.
\end{equation}
If one therefore solves Eq. (\ref{jt}) between the initial and final states in both angular momentum as $J\in[J_0,~J_C]$ and time as $t\in[0,~\sqrt{2\Omega_0/\epsilon}]$, one finds rather interestingly that $J_C\simeq 0.794J_0$. Due to the non-conservative mechanics imposed, angular momentum is lost and thus reduced over time.

In addition, if one equates $t_C=\sqrt{2\Omega_0/\epsilon}$ to $t_C$ via Eq. (\ref{ins}) under $\beta\sim0.1$ and $\widetilde{P}\approx6$, then one finds that
\begin{equation}
\kappa^2\left(m_1m_2M\right)^{2/3}=\frac{2\Omega_0}{\epsilon},
\end{equation}
where $\kappa\simeq2.214\times10^{-29}~\mathrm{s/kg}$ is the corresponding coefficient implied in Eq. (\ref{ins}). Therefore, in terms of $\Omega_0$ (the initial-state circular orbit frequency), one yields $\epsilon\propto \Omega_0(m_1m_2M)^{-2/3}\sim\Omega_0M^{-2/3}$.

\section{Concluding Statements} \label{concl}

In this work, we analyzed the CBB mass-shell model with necessary non-conservative mechanics. This has allowed one to obtain the mass-shell-modelled waveform profiles given by Eqs. (\ref{hp}) and (\ref{hc}), which readily include dynamic information of the CBB orbital (mass-shell axial) frequency and its rates of change in the time domain. In addition to these waveforms, and their symmetric relation with phenomenological models via Eq. (\ref{symms}), we explored how the mass-shell model of a remapped CBB gives better analytical intuition of early-inspiral information: from an order-of-magnitude estimate of the coalescence timelapse from this phase (Eq. [\ref{ins}]) to minimal GW frequency, which is discernable via Eqs. (\ref{l0}) and (\ref{tc}). Near-merger information can also be discerned from this mass-shell model, such as the previously analyzed GW energy at merger (Eq. (\ref{energy}), c.f. Ref. \cite{MacKay:2025uyg}) and anticipated maximum GW frequency (Eq. [\ref{freq}]) for total masses $M>10M_\odot$. 

It is a sincere hope that this work and the previous works \cite{MacKay:2024qxj, MacKay:2025uyg} become another means to model waveforms and GW energetics analytically. This is more so for next-generation detectors and their respective collaborations, whose research statement is to e.g. capture early-stage and low-frequency GWs from CBBs. Future endeavours from this work span into a complete numerical implementation of Eqs. (\ref{hp}) and (\ref{hc}), as well as to explore model extensions -- both analytical and numerical -- to include internal post-Newtonian corrections, as discussed in Ref. \cite{MacKay:2025uyg}, and external, environmental effects such as circumbinary baryonic and/or dark matter, as discussed in Ref.  \cite{MacKay:2024qxj}.

\begin{acknowledgments}
I thank Tim Dietrich for valuable conversations on non-conservative mechanisms in current and state-of-the-art waveform modelling, and for inspiring how this may be accounted for in this novel CBB mass-shell approach.
\end{acknowledgments}





\end{document}